\documentclass{aa}
\usepackage{graphics}
\usepackage{psfig}
\def\h2{H{\small II}}
\begin{document}
%-----------------------------------------------------------------
\title{Pox 186: an ultracompact galaxy with dominant ionized gas emission}
%
%\subtitle{A Clue to Dwarf Galaxy Evolution?}
%
\author{N. G. Guseva \inst{1}
\and P. Papaderos \inst{2}
\and Y. I. Izotov \inst{1}
\and K. G. Noeske\inst{2,}\inst{3}
\and K. J. Fricke \inst{2}}
%\and T. X. Thuan\inst{4}}
%
\offprints{N.G. Guseva, guseva@mao.kiev.ua}
\institute{      Main Astronomical Observatory,
                 Ukrainian National Academy of Sciences,
                 Zabolotnoho 27, Kyiv 03680,  Ukraine
\and
                 Universit\"ats--Sternwarte, Geismarlandstra\ss e 11,
                 D--37083 G\"ottingen, Germany
\and
                 University of California, 1156 High St., Santa Cruz,
                 CA 95064, USA
%\and
%                 Astronomy Department, University of Virginia, 
%                 Charlottesville, VA 22903, USA
}

\date{Received \hskip 2cm; Accepted}
%\date{Received(00.00.0000)/Accepted(00.00.0000)}
%\titlerunning{A spectroscopic and photometric study of SBS 1415+437}
%\authorrunning{N. G. Guseva et al.}
%\maketitle

% ======================================================================================
\abstract{We present a ground-based optical spectroscopic and {\sl HST}
$U$, $V$, $I$ 
photometric study of the blue compact dwarf (BCD) galaxy Pox 186. 
It is found that the emission of the low-surface brightness (LSB) component 
in Pox 186 at radii $\la$ 3\arcsec\ ($\la$ 270 pc in linear scale) is mainly 
gaseous in origin. We detect H$\alpha$ emission out to radii
as large as 6\arcsec. At radii $\ga$3\arcsec\ the light of the LSB 
component is contaminated by the emission
of background galaxies complicating the study of the outermost regions.
The surface brightness distribution in the LSB component can be
approximated by an exponential law with a scale length
$\alpha$ $\la$ 120 pc. This places Pox 186  
among the most compact dwarf galaxies known. The derived $\alpha$ 
is likely to be an upper limit to the scale length 
of the LSB component because of 
the strong contribution of the gaseous emission.
The oxygen abundance in the bright H {\sc ii} region derived from
the 4.5m Multiple Mirror Telescope
(MMT)\thanks{The MMT Observatory is a joint facility of the 
Smithsonian Institution and the University of Arizona.} 
and 3.6m ESO telescope\thanks{Based on observations
collected at the European Southern Observatory, Chile, ESO program 
71.B-0032(A).} spectra are 
12 + log(O/H) = 7.76 $\pm$ 0.02 and 7.74 $\pm$ 0.01 
($\sim$$Z_\odot$/15)\thanks{12+log(O/H)$_{\odot}$ = 8.92
(Anders \& Grevesse \cite{Anders89}).}, 
respectively, in accordance with previous determinations.
The helium mass fractions found in this region are $Y$ = 0.248$\pm$0.009 (MMT)
and $Y$ = 0.248$\pm$0.004 (3.6m) suggesting a high 
primordial helium abundance. 
%We find that the emission line flux ratios and 
%heavy element to oxygen abundance ratios are best reproduced if 
%the H {\sc ii} region in Pox 186 is density-bounded. However, some neutral
%gas is likely to be present in Pox 186 because weak 
%[O {\sc i}] $\lambda$6300\AA\ emission has been detected in its spectrum.
}

\maketitle
%---------------------------------------------------------------------------------------------

\markboth {N. G. Guseva et al.}{Pox 186: an ultracompact galaxy with dominant ionized gas emission}
%----------------------------------------------------------------------------------------------
\section {Introduction}
\label{intro}

The blue compact dwarf (BCD) galaxy Pox 186 $\equiv$ PGC 046982 was 
classified by Kunth, Sargent \& Kowal (\cite{K81}) as an ultracompact
galaxy at redshift $z$ = 0.003903 with an almost stellar appearance and 
strong emission lines in its spectrum.
It was spectroscopically studied by Kunth \& Sargent (\cite{K83}). 
They derived an oxygen abundance 12 + log(O/H) = 7.72 $\pm$ 0.02.
Later, Kunth, Maurogordato \&
Vigroux (\cite{KMV88}) presented the results of a $B$, $R$, $I$ photometric
study of Pox 186. They found that the galaxy is very compact with a
diameter of only $\sim$ 350 pc at a surface brightness
$\mu$ = 25 $R$ mag arcsec$^{-2}$. Since no H {\sc i} emission was detected 
with the Nan\c{c}ay radio telescope, Kunth et al. (\cite{KMV88}) put
an upper limit of 2$\times$10$^7$ $M_\odot$ on the neutral hydrogen mass.

Using ground-based $R$ and $I$ photometric data, Doublier et al.
(\cite{D00}) found that the central star-forming region 
of Pox 186 consists of three
compact super-star clusters (SSCs). They measured a
relatively red $R-I$ colour in the extended component at radii $>$ 2\arcsec\ 
suggesting that the 
stellar populations in the low-surface-brightness (LSB) component of Pox 186 
are old.

Corbin \& Vacca (\cite{V02}) obtained {\sl HST} $U$, $V$, $I$ photometric
and UV spectroscopic data as well as ground-based optical spectra. They do not 
find multiple super-star clusters, contrary to Doublier et al. (\cite{D00}).
Corbin \& Vacca (\cite{V02})  
also concluded that the red $V-I$ colour of the extended component is mostly 
stellar, dominated by K and M stars.
Based on the irregular morphology and the presence
of a tail in the central part of the galaxy they suggested that this system 
is in the stage of formation after merging of two smaller galaxies.

Pox 186 with its
small linear extent belongs to the rare class of very compact dwarf galaxies
and may be considered an example of the building blocks from which larger
galaxies may have formed. This strengthens the motivation for more detailed
studies of this object. The conclusions obtained from previous papers on the 
properties and the evolutionary status of Pox 186 remain controversial.
In the most recent papers Doublier et al. (\cite{D00}) and 
Corbin \& Vacca (\cite{V02}) argue, based on broad-band photometric 
data alone, that the ionized gas emission is not
important in the LSB component, while Gil de Paz, Madore \& Pevunova 
(\cite{GMP03}) have clearly detected nebular emission at radii 
$\ga$ 2\arcsec\ in the H$\alpha$ image of Pox 186.

In this paper we use new spectroscopic observations of Pox 186
and re-analyze archival {\sl HST} images in order to better understand
morphological and physical properties of this galaxy. Observations and
data reduction are described in Sect. 2. We present the results in Sect. 3. 
Our main conclusions are summarized in Sect. 4.

%***********************************************************************
%                 Figure 1
%***********************************************************************
\begin{figure}[hbtp]%[tbh]
\begin{picture}(8.5,8.5)
\put(0,0){{\psfig{figure=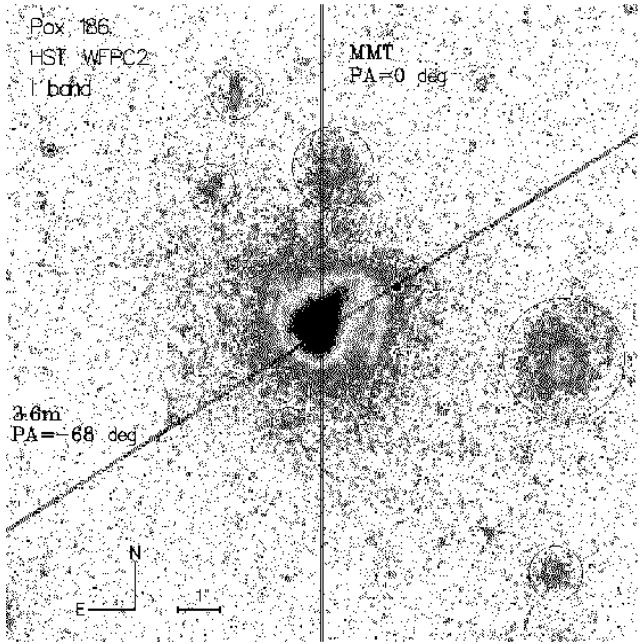,height=8.5cm,angle=0.,clip=}}}
\end{picture}
  \caption{{{\sl HST} WFPC2 image of Pox 186 in $I$ 
with red background galaxies depicted with circles and a probable star
marked with a cross. The horizontal bar is 
1\arcsec, corresponding to a linear scale of 90 pc. North is up and east 
to the left. The placement of the long slit during the MMT and 3.6m telescope 
observations at position angles of respectively 0$\degr$ and --68$\degr$ is 
shown.}
}
    \label{f:ima1}
\end{figure}
% *********************************************************************

% =======================================
\section{Observations and data reduction \label{obs}}

%*************************************************************
%             Fig.2
%*************************************************************
\begin{figure}%[hbtp]%[tbh]
%\vspace{7.cm}
   \hspace*{0.0cm}\psfig{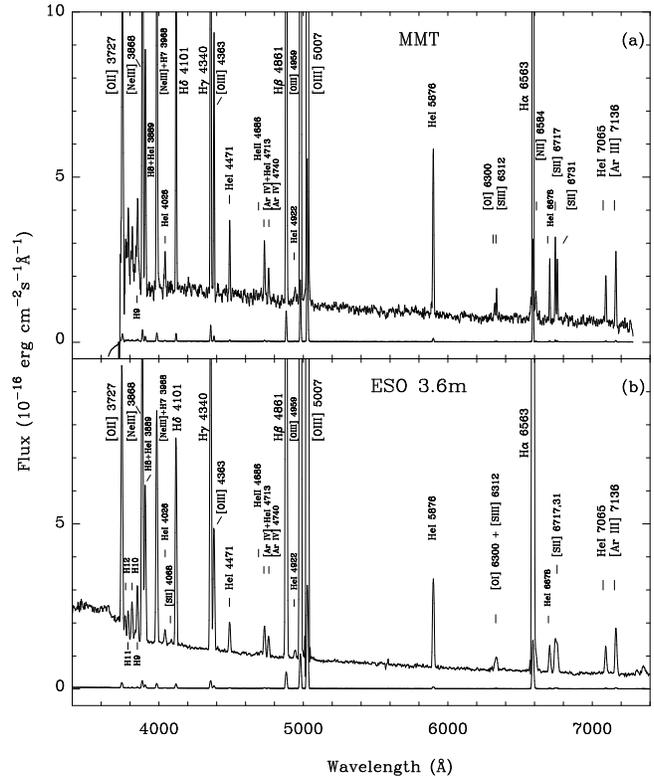}
    \caption{The MMT and 3.6m telescope spectra of the 
bright H {\sc ii} region with the emission lines labeled. The lower spectra
in (a) and (b) are the observed spectra downscaled by a factors of 50.
      }
    \label{f:bright}
\end{figure}
%*************************************************************

\subsection{Spectroscopy}
%-------------------------------------------

 The 4.5m  Multiple Mirror Telescope (MMT) long-slit spectroscopic 
observations over a wavelength range 3650 -- 7500\AA\ were carried out on 
20 April, 1994.
The blue channel spectrograph with a
3072~$\times$~1024 CCD detector in conjunction with a 
2\arcsec~$\times$~180\arcsec\ slit, and a grating with 500 grooves mm$^{-1}$
and a second-order blocking filter L-37 were used.
The spatial scale along the slit was 0\farcs3~pixel$^{-1}$. The spectrum
was binned in the spatial direction resulting in a spatial resolution of
0\farcs6. The spectral resolution was $\sim$ 7~\AA\ (FWHM). 
The  slit was oriented at position angle P.A. = 0$\degr$
and centered on the bright H {\sc ii} region ({Fig. \ref{f:ima1}). 
The spectrum was obtained at an airmass of 1.64 and a seeing of 
$\sim$ 1\arcsec. 
The total exposure time of 2400 s was broken 
into three subexposures of 600 s and 2$\times$900 s.
No correction for atmospheric refraction was applied because
the observations were done at a position angle close to the parallactic one.
Two spectrophotometric standard stars were 
observed for flux calibration.
For wavelength calibration spectra of a He-Ne-Ar comparison lamp were obtained 
after each exposure.

The observations with the 3.6-m ESO telescope (La Silla) were obtained on 
25 April, 2003. We used a grism $\#11$ ($\lambda$$\lambda$3400--7400)
and no second-order blocking filter. 
The long slit with the width of 1\arcsec\ was centered 
on the brightest part of the H {\sc ii} region at position angle
P.A. = --68$\degr$.  
The seeing was $\sim$ 0\farcs65.
The spectra were obtained at airmass 1.113. The total 
exposure time of 3600 s was split up into three subexposures.
The spectra were binned along the spatial and dispersion axes resulting in 
a spatial scale along the slit of 0\farcs314 pixel$^{-1}$, and a
spectral resolution of $\sim$13.2~\AA\ (FWHM). 
No correction for atmospheric refraction was made because of 
the small airmass during the observations.
Three spectrophotometric standard stars were observed for flux calibration.

The data reduction was made with the IRAF\footnote{IRAF is 
the Image Reduction and Analysis Facility distributed by the 
National Optical Astronomy Observatory, which is operated by the 
Association of Universities for Research in Astronomy (AURA) under 
cooperative agreement with the National Science Foundation (NSF).}
software package. This includes  bias--subtraction, 
flat--field correction, cosmic-ray removal, wavelength calibration, 
night sky background subtraction, correction for atmospheric extinction and 
absolute flux calibration of the two--dimensional spectrum.

One-dimensional spectra of the bright H {\sc ii} region 
 were extracted within apertures of 
2\arcsec\ $\times$ 6\arcsec\ and 1\arcsec\ $\times$ 3\farcs2
from the MMT and 3.6m data, respectively, 
and are shown in Fig.~\ref{f:bright}. 
The strong increase of the continuum shortward
$\sim$ 3660\AA\ in the 3.6m spectrum is due to the Balmer jump in the ionized
gas emission, indicating that the contribution of the ionized gas to the total
emission is very large.

The presence of numerous strong emission lines in the spectra of Pox 186
allows the reliable determination of the BCD redshift. The redshift of 
Pox 186 as derived from the observed wavelengths of the 17 brightest 
emission lines in the MMT spectrum and of the 27 brightest emission lines in 
the 3.6m spectrum is respectively $z$ = 0.00413$\pm$0.00005 and  
$z$ = 0.00413$\pm$0.00010. After correction of $z$ for 
the motion relative to the Local Group centroid and the Virgocentric 
flow (Kraan-Korteweg \cite{K86}) we derive the distance to Pox 186 as 
18.5 Mpc, assuming a Hubble constant of 75 km s$^{-1}$ Mpc$^{-1}$. 
At this distance 1\arcsec\ corresponds to a linear extent of 90 pc. 
%assuming Hubble constant of $H_0$ = 75 km s$^{-1}$ Mpc$^{-1}$.

\subsection{Photometry}

We use the broad-band {\sl HST} WFPC2 images of Pox 186 in
F336W (2$\times$700 s), F555W (600 s) and
F814W (2$\times$2700 s) from the {\sl HST} data archive ({\sl HST} 
Proposal ID:  8333, PI: M. Corbin). 
The images were reduced and calibrated by the pipeline package at the 
Space Telescope Science Institute (STScI) and further processed using
ESO MIDAS\footnote{Munich Image Data Analysis System, provided by the
European Southern Observatory (ESO).}.
In all bands the object was centered on the PC chip, yielding a 
scale of 0\farcs 046 pixel$^{-1}$. We combined the images in F336W and F814W
and removed the cosmic rays using the IRAF/STSDAS routine CRREJ.
Only a single F555W image was obtained. Therefore, the cosmic rays in this 
image were removed manually. We converted instrumental F336W, F555W, F814W 
magnitudes to the standard $U$, $V$, $I$ magnitudes following the 
prescriptions by Holtzman et al. (\cite{H95}).

All images were corrected for Galactic extinction assuming
$A_B$ = 0.201 mag, based on Schlegel, Finkbeiner \& Davis (\cite{S98}; 
see NASA Extragalactic Database (NED)).

%**************************************************************
%     Table 1
%**************************************************************

\begin{table*}[tbh]
%     \centering{
\caption{Fluxes and equivalent widths of emission lines in 
the bright H {\sc ii} region.
%Observed ($F$($\lambda$)) and extinction-corrected  
%($I$($\lambda$)) fluxes and 
% equivalent widths ($EW$) of emission lines in H {\sc ii} region.
}
\label{t:Intens}
\begin{tabular}{lccrcccr} \hline \hline
  &\multicolumn{3}{c}{MMT}&&\multicolumn{3}{c}{3.6m} \\ \cline{2-4} \cline{6-8}
$\lambda_{0}$(\AA) Ion                  &$F$($\lambda$)/$F$(H$\beta$)$^{\rm a}$
&$I$($\lambda$)/$I$(H$\beta$)$^{\rm b}$&$EW$(\AA)$^{\rm c}$&&$F$($\lambda$)/$F$(H$\beta$)$^{\rm a}$&$I$($\lambda$)/$I$(H$\beta$)$^{\rm b}$ 
&$EW$(\AA)$^{\rm c}$  \\ \hline
3727\ [O {\sc ii}]           & 0.279 $\pm$0.075 & 0.279 $\pm$0.076 &88.6 $\pm$22.0 && 0.302$\pm$0.002 & 0.338$\pm$0.002 & 54.6$\pm$0.3 \\
3750\ H12                    &       ...        &       ...        &\multicolumn {1}{c}{...} && 0.014$\pm$0.001 & 0.023$\pm$0.003 & 2.8$\pm$0.3 \\
3771\ H11                    &       ...        &       ...        &\multicolumn {1}{c}{...} && 0.027$\pm$0.001 & 0.037$\pm$0.002 & 6.0$\pm$0.3 \\
3798\ H10                    &       ...        &       ...        &\multicolumn {1}{c}{...} && 0.041$\pm$0.001 & 0.052$\pm$0.002 & 9.5$\pm$0.3 \\
3835\ H9                     & 0.052 $\pm$0.013 & 0.073 $\pm$0.021 &12.1 $\pm$2.3  && 0.062$\pm$0.001 & 0.075$\pm$0.002 & 14.8$\pm$0.3 \\
3868\ [Ne {\sc iii}]         & 0.420 $\pm$0.014 & 0.419 $\pm$0.014 &120.6$\pm$2.5  && 0.421$\pm$0.002 & 0.464$\pm$0.002 &101.9$\pm$0.4 \\
3889\ H8 + He {\sc i}           & 0.173 $\pm$0.010 & 0.190 $\pm$0.014 &47.8$\pm$2.2  && 0.184$\pm$0.001 & 0.208$\pm$0.002 & 45.2$\pm$0.3 \\
3968\ [Ne {\sc iii}] + H7       & 0.287 $\pm$0.007 & 0.305 $\pm$0.012 &73.5 $\pm$1.3  && 0.271$\pm$0.001 & 0.302$\pm$0.002 & 68.4$\pm$0.3 \\
4026\ He {\sc i}             & 0.022 $\pm$0.004 & 0.022 $\pm$0.004 & 6.0 $\pm$1.0  && 0.018$\pm$0.001 & 0.020$\pm$0.001 &  4.7$\pm$0.2 \\
4068\ [S {\sc ii}]           &       ...        &       ...        &\multicolumn {1}{c}{...} && 0.008$\pm$0.002 & 0.009$\pm$0.002 & 2.2$\pm$0.4 \\
4101\ H$\delta$              & 0.228 $\pm$0.005 & 0.245 $\pm$0.010 &60.0 $\pm$0.9  && 0.239$\pm$0.001 & 0.262$\pm$0.002 & 64.4$\pm$0.3 \\
4340\ H$\gamma$              & 0.455 $\pm$0.008 & 0.468 $\pm$0.012 &126.8$\pm$1.0  && 0.447$\pm$0.002 & 0.473$\pm$0.002 &131.3$\pm$0.4 \\
4363\ [O {\sc iii}]          & 0.144 $\pm$0.004 & 0.143 $\pm$0.004 &40.3 $\pm$0.9  && 0.148$\pm$0.001 & 0.154$\pm$0.001 & 43.6$\pm$0.3 \\
4471\ He {\sc i}             & 0.040 $\pm$0.003 & 0.040 $\pm$0.003 &11.1 $\pm$0.8  && 0.037$\pm$0.001 & 0.038$\pm$0.001 & 11.4$\pm$0.2 \\
4686\ He {\sc ii}            & 0.005 $\pm$0.002 & 0.005 $\pm$0.002 & 1.5 $\pm$0.6  && 0.011$\pm$0.001 & 0.011$\pm$0.001 &  3.9$\pm$0.5 \\
4713\ [Ar {\sc iv}] + He {\sc i} & 0.037 $\pm$0.003 & 0.037 $\pm$0.003 &11.5 $\pm$0.8  && 0.041$\pm$0.001 & 0.042$\pm$0.001 & 14.9$\pm$0.3 \\
4740\ [Ar {\sc iv}]          & 0.019 $\pm$0.003 & 0.019 $\pm$0.003 & 6.0 $\pm$0.8  && 0.026$\pm$0.001 & 0.026$\pm$0.001 &  9.5$\pm$0.3 \\
4861\ H$\beta$               & 1.000 $\pm$0.015 & 1.000 $\pm$0.017 &334.9$\pm$1.1  && 1.000$\pm$0.002 & 1.000$\pm$0.002 &375.0$\pm$0.6 \\
4922\ He {\sc i}             & 0.012 $\pm$0.002 & 0.012 $\pm$0.002 & 4.4 $\pm$0.6  && 0.013$\pm$0.001 & 0.013$\pm$0.001 &  5.0$\pm$0.4 \\
4959\ [O {\sc iii}]          & 2.086 $\pm$0.031 & 2.054 $\pm$0.031 &726.7$\pm$1.3  && 2.094$\pm$0.004 & 2.069$\pm$0.004 &817.9$\pm$0.9 \\
5007\ [O {\sc iii}]          & 6.212 $\pm$0.091 & 6.114 $\pm$0.091 &2208.0$\pm$2.0 && 6.295$\pm$0.011 & 6.192$\pm$0.011 &2497.0$\pm$1.5 \\
5876\ He {\sc i}             & 0.112 $\pm$0.003 & 0.109 $\pm$0.003 &56.6 $\pm$1.2  && 0.111$\pm$0.001 & 0.102$\pm$0.001 & 58.3$\pm$0.6 \\
6300\ [O {\sc i}]            & 0.010 $\pm$0.002 & 0.010 $\pm$0.002 & 5.5 $\pm$1.2  && 0.009$\pm$0.001 & 0.008$\pm$0.001 &  5.4$\pm$0.4 \\
6312\ [S {\sc iii}]          & 0.015 $\pm$0.002 & 0.015 $\pm$0.002 & 8.4 $\pm$1.1  && 0.014$\pm$0.001 & 0.013$\pm$0.001 &  8.5$\pm$0.4 \\
6563\ H$\alpha$              & 2.850 $\pm$0.042 & 2.775 $\pm$0.045 &1620.0$\pm$2.4 && 3.159$\pm$0.006 & 2.776$\pm$0.006 &2075.0$\pm$1.9 \\
6584\ [N {\sc ii}]           & 0.017 $\pm$0.002 & 0.016 $\pm$0.002 & 9.9 $\pm$1.3  &&        ...      &        ...      &\multicolumn {1}{c}{...} \\
6678\ He {\sc i}             & 0.029 $\pm$0.002 & 0.028 $\pm$0.002 &16.8 $\pm$1.2  && 0.033$\pm$0.001 & 0.029$\pm$0.001 & 22.7$\pm$0.6 \\
6717\ [S {\sc ii}]           & 0.045 $\pm$0.003 & 0.044 $\pm$0.003 &27.7 $\pm$1.4  && 0.032$\pm$0.001 & 0.028$\pm$0.001 & 22.6$\pm$0.4 \\
6731\ [S {\sc ii}]           & 0.033 $\pm$0.003 & 0.032 $\pm$0.003 &19.6 $\pm$1.5  && 0.028$\pm$0.001 & 0.025$\pm$0.001 & 19.9$\pm$0.4 \\
7065\ He {\sc i}             & 0.031 $\pm$0.003 & 0.030 $\pm$0.003 &18.4 $\pm$2.2  && 0.034$\pm$0.001 & 0.029$\pm$0.001 & 24.3$\pm$0.7 \\
7136\ [Ar {\sc iii}]         & 0.051 $\pm$0.004 & 0.049 $\pm$0.004 &39.3 $\pm$3.9  && 0.063$\pm$0.001 & 0.054$\pm$0.001 & 46.7$\pm$0.8 \\
                     & & & && & & \\
$C$(H$\beta$)\ dex             &\multicolumn {3}{c}{0.020$\pm$0.019} &&\multicolumn {3}{c}{0.165$\pm$0.002}\\
$F$(H$\beta$)$^{\rm a,d}$        &\multicolumn {3}{c}{4.08$\pm$0.01}   &&\multicolumn {3}{c}{3.48$\pm$0.01} \\
$EW$(abs)~\AA$^{\rm e}$                  &\multicolumn {3}{c}{4.8$\pm$2.3}     &&\multicolumn {3}{c}{1.4$\pm$0.3}  \\
\hline%\hline
\end{tabular}

$^{\rm a}$observed fluxes. \\ 
$^{\rm b}$fluxes corrected for extinction and underlying stellar absorption. \\
$^{\rm c}$measured equivalent widths of the emission lines. \\
%$^{\rm c}$equivalent widths in \AA. \\
$^{\rm d}$in units 10$^{-14}$\ erg\ s$^{-1}$cm$^{-2}$. \\
$^{\rm e}$calculated equivalent widths of underlying hydrogen 
absorption lines.
%}
\end{table*}
%************************************************************************

\section{Results \label{Sect:results}}

\subsection{Element abundances in the central star-forming region 
\label{Sect:burst}}

 In this section we derive the element abundances in 
the bright H {\sc ii} region of Pox 186
and study 
%with
% the CLOUDY code (Ferland \cite{F96}; Ferland et al. \cite{F98}) 
its physical properties. 
Spectra used for the abundance determination are shown in Fig.~\ref{f:bright}.

%____________________________________________________________________
%\subsubsection{Chemical abundances \label{chem}}
%--------------------------------------------------------------------

The fluxes and equivalent widths of the emission lines with the errors
have been measured by fitting Gaussians to the line profiles. For
this we use the IRAF routine SPLOT. The errors have been propagated through the
determination of the element abundances.
Following the procedure detailed in Izotov et al. (\cite{ITL94,ITL97})
we corrected the observed emission line fluxes for the interstellar 
extinction and underlying hydrogen
stellar absorption lines using the observed Balmer decrement of hydrogen 
emission lines. The correction was done by minimizing the deviations
of corrected hydrogen emission line flux ratios from the theoretical ones.
For this the H9 and H$\delta$ -- H$\alpha$ emission lines were used to correct
the MMT spectrum, and the H12 -- H9, H$\delta$ -- H$\alpha$ emission lines
were used to correct the 3.6m spectrum. The H7 and H8 lines were not used
because they are blended with the other emission lines. The observed 
($F$($\lambda$)) and corrected 
($I$($\lambda$)) emission line fluxes relative to the H$\beta$ emission line 
fluxes, their equivalent widths $EW$, the extinction coefficients 
$C$(H$\beta$), the observed fluxes of the H$\beta$ emission line, and the 
equivalent widths of the hydrogen absorption lines
for the bright H {\sc ii} region are shown in Table \ref{t:Intens} 
for the MMT and 3.6m data, respectively. 

In general, the agreement between
the corrected emission line fluxes and their equivalent widths 
for the two observations is good.
Note the very high equivalent widths of the H$\beta$ and 
H$\alpha$ emission lines, implying a young
age of the starburst in the central part of Pox 186. 
The extinction coefficients $C$(H$\beta$) = 0.02 $\pm$ 0.02 (MMT) and 
0.16 $\pm$ 0.00 (3.6m) are in good agreement with the typical values of
BCDs. We assume that $C$(H$\beta$) derived from
the 3.6m spectrum is more reliable because of its higher
signal-to-noise ratio and the lower airmass during the observations. 
Our derived $C$(H$\beta$) differs from the value $C$(H$\beta$) = 0.41 
($E$($B-V$) = 0.28) by Corbin \& Vacca (\cite{V02}). Note that 
the H$\beta$ flux in both of our spectra is $\sim$ 2 times larger than that 
measured by Corbin \& Vacca (\cite{V02}) within a larger aperture. However,
our values are consistent with the H$\alpha$ flux derived by
Gil de Paz et al. (\cite{GMP03}) from narrow-band imaging of Pox 186.

The electron temperature $T_{\rm e}$, electron number density $N_{\rm e}$,
ionic and total heavy element abundances were derived 
following Izotov et al. (\cite{ITL94,ITL97}) and Thuan et al. (\cite{til95}).
Because the resolution of the 3.6m spectrum is low the [S {\sc ii}]
$\lambda$6717 and $\lambda$6731 lines 
are blended. Therefore, the number density 
$N_{\rm e}$(S {\sc ii}) derived from this spectrum by deblending the 
[S {\sc ii}] doublet is not certain.
The same is valid for the sulfur abundance which is derived from the 
[S {\sc ii}]
$\lambda$6717+6731 and [S {\sc iii}] $\lambda$6312 emission lines.
The auroral [S {\sc iii}] $\lambda$6312 emission line in the 3.6m spectrum
is blended with the [O {\sc i}] $\lambda$6300 emission line and its flux
was derived after a deblending procedure. 
The flux of the [Ar {\sc iii}]
$\lambda$7135 emission line and hence the Ar abundance derived from the
3.6m spectrum may be affected by contamination with the second order because
no blocking filter was used.
Finally, no nitrogen abundance was derived from the 3.6m spectrum because
the weak [N {\sc ii}] $\lambda$6584 emission line is blended with strong 
H$\alpha$.

The electron temperatures $T_{\rm e}$(O {\sc iii}), 
$T_{\rm e}$(S {\sc iii}) $\equiv$ $T_{\rm e}$(Ar {\sc iii}),
$T_{\rm e}$(O {\sc ii}) for the high-, intermediate- and low-ionization 
regions respectively, the electron number 
densities $N_{\rm e}$(S {\sc ii}), ionization correction factors (ICF), 
ionic and total heavy element abundances are shown in Table \ref{t:Chem}.
The oxygen abundance 12 + log(O/H) = 7.76 $\pm$ 0.02
(MMT) and 7.74 $\pm$ 0.01 (3.6m)
for the bright H {\sc ii} region are within the errors
in agreement with 12 + log(O/H) = 7.72 $\pm$ 0.02 derived
by Kunth \& Sargent (\cite{K83}) and $>$ 7.73 derived by 
Corbin \& Vacca (\cite{V02}).

The five brightest He {\sc i} $\lambda$3889, $\lambda$4471, $\lambda$5876, 
$\lambda$6678, $\lambda$7065 emission lines were used for determining the 
He abundance. Their fluxes were corrected for collisional and fluorescent 
enhancements according to Izotov et al. (\cite{ITL94,ITL97}). 
This was made by minimizing the deviations of the 
corrected He {\sc i} line flux ratios from the recombination ratios, varying 
the electron number density in He$^+$ zone and optical depth in the 
He {\sc i} $\lambda$3889 transition. 
Singly ionized helium abundances He$^+$/H$^+$ derived from the corrected 
He {\sc i} $\lambda$4471, $\lambda$5876, $\lambda$6678 line fluxes and their 
weighted mean are shown in Table \ref{t:Chem}. A small contribution of 
He$^{+2}$ was added, because He {\sc ii} $\lambda$4686 is present.
The mean $^4$He mass
fractions $Y$ = 0.248$\pm$0.009 (MMT) and 0.248$\pm$0.004 (3.6m) 
(Table \ref{t:Chem})
are consistent with the value $Y$ = 0.245$\pm$0.012 derived for Pox 186  by  
Kunth \& Sargent (\cite{K83}). They are also consistent with the primordial 
$^4$He mass fraction $Y_{\rm p}$ = 0.244 $\pm$ 0.002, derived by extrapolating
the $Y$ vs O/H linear regression to O/H = 0 (Izotov \& Thuan \cite{IT98}), 
or with $Y_{\rm p}$ = 0.245 $\pm$ 0.002 derived from the spectroscopic 
observations of the two most metal-deficient BCDs I Zw 18 and SBS 0335--052 
(Izotov et al. \cite{I99}). 

 The N/O, Ne/O, S/O and Ar/O abundance ratios vs oxygen abundance 
12 + log(O/H) in Pox 186 are shown in Fig.~\ref{fig:heavy} by the filled
triangles and filled squares, respectively, for the MMT and 3.6m 
observations. For comparison, by the open circles are shown the BCDs from 
Izotov \& Thuan (\cite{IT99}). 
We find a rather high value of log(N/O) in Pox 186 for its oxygen 
abundance (Fig.~\ref{fig:heavy}) which is however in agreement 
with log(N/O) = --1.38 derived by Kunth \& Sargent (\cite{K83}) 
and with the value of --1.42 recomputed by Kobulnicky \& Skillman 
(\cite{Kob96}) from the Kunth \& Sargent data.

The most striking
feature of Fig. \ref{fig:heavy} is that the S/O abundance ratio derived from
the MMT spectrum is $\sim$ 2 times higher than the mean value for BCDs.
On the other hand the sulfur abundance derived from the 3.6m spectrum 
is lower. However, the [S {\sc iii}] $\lambda$6312 emission line in the 3.6m
spectrum is blended with the [O {\sc i}] $\lambda$6300 emission line.
This makes the sulfur abundance derived from the 3.6m spectrum less 
reliable.
Sulfur, similar to oxygen, neon and argon, is produced by massive stars.
The Ne/O abundance ratio in Pox 186 does not show a deviation from 
the mean BCD value and the Ar/O abundance ratio is slightly above the mean. 
Therefore, no significant deviation is expected for the S/O abundance ratio. 
Thus, we conclude that the apparently high S/O abundance ratio in Pox 186 
is probably not due to nucleosynthetic processes.
%but by the structure of the 
%H {\sc ii} region.

%***********************************************************
%    Table 2
%***********************************************************

\begin{table}[tbh]
\caption{Element abundances in the bright H {\sc ii} region.}
\label{t:Chem}
\begin{tabular}{lccc} \hline \hline
Value                               &  MMT      &&   3.6m  \\ \hline
$T_{\rm e}$(O {\sc iii})(K)               &16390$\pm$240  && 16940$\pm$60 \\
$T_{\rm e}$(O {\sc ii})(K)                &14390$\pm$200  && 14610$\pm$50 \\
$T_{\rm e}$(Ar {\sc iii})(K)              &15300$\pm$200  && 15760$\pm$50 \\
$T_{\rm e}$(S {\sc iii})(K)               &15300$\pm$200  && 15760$\pm$50 \\
$N_{\rm e}$(S {\sc ii})(cm$^{-3}$)        & 70$\pm$110    && 350$\pm$60 \\ \\
%$N_{\rm e}$(He {\sc ii})(cm$^{-3}$)             &    ..$\pm$..  
%$\tau$($\lambda$3889)               &       0.0     
O$^+$/H$^+$($\times$10$^5$)         &0.276$\pm$0.076 && 0.329$\pm$0.010 \\
O$^{+2}$/H$^+$($\times$10$^5$)      &5.387$\pm$0.205 && 5.057$\pm$0.137 \\
O$^{+3}$/H$^+$($\times$10$^6$)      &0.295$\pm$0.130 && 0.064$\pm$0.008 \\
O/H($\times$10$^5$)                 &5.693$\pm$0.219 && 5.449$\pm$0.138 \\
12 + log(O/H)                       &7.76$\pm$0.02   && 7.74$\pm$0.01 \\ \\
%---------------------------------------------------------------------------
N$^{+}$/H$^+$($\times$10$^7$)       &1.305$\pm$0.124 &&\multicolumn {1}{c}{...} \\
ICF(N)$^{\rm a}$                          &20.62~~~~~~~~~~&&\multicolumn {1}{c}{...} \\
log(N/O)                             &--1.33$\pm$0.04~~~~~&&\multicolumn {1}{c}{...} \\ \\
%---------------------------------------------------------------------------
Ne$^{+2}$/H$^+$($\times$10$^5$)     &0.796$\pm$0.039 && 0.808$\pm$0.023 \\
ICF(Ne)$^{\rm a}$                         &1.06~~~~~~~~~~~~&&1.08~~~~~~~~~~~~\\
log(Ne/O)                            &--0.83$\pm$0.03~~~~~&&--0.80$\pm$0.02~~~~~\\ \\
%---------------------------------------------------------------------------
S$^+$/H$^+$($\times$10$^7$)         &0.809$\pm$0.043 && 0.568$\pm$0.014 \\
S$^{+2}$/H$^+$($\times$10$^7$)      &7.081$\pm$1.108 && 5.675$\pm$0.324 \\
ICF(S)$^{\rm a}$                          &4.31~~~~~~~~~~~~&&3.60~~~~~~~~~~~~\\
log(S/O)                             &--1.22$\pm$0.06~~~~~&&--1.38$\pm$0.03~~~~~\\ \\
%---------------------------------------------------------------------------
Ar$^{+2}$/H$^+$($\times$10$^7$)      &1.723$\pm$0.145 && 1.786$\pm$0.047 \\
Ar$^{+3}$/H$^+$($\times$10$^7$)      &2.234$\pm$0.321 && 2.923$\pm$0.112 \\
ICF(Ar)$^{\rm a}$                          &1.01~~~~~~~~~~~~&&1.00~~~~~~~~~~~~\\
log(Ar/O)                             &--2.16$\pm$0.04~~~~~&&--2.06$\pm$0.02~~~~~\\ \\
%----------------------------------------------------------------------------
% for He abundance it is used the data from Smith, 1996 for 3.6m data
He$^+$/H$^+$($\lambda$4471)         &0.081$\pm$0.006 && 0.080$\pm$0.002 \\
He$^+$/H$^+$($\lambda$5876)         &0.084$\pm$0.004 && 0.082$\pm$0.002 \\
He$^+$/H$^+$($\lambda$6678)         &0.080$\pm$0.006 && 0.084$\pm$0.003 \\
He$^+$/H$^+$  (mean)                &0.082$\pm$0.003 && 0.082$\pm$0.001 \\
He$^{+2}$/H$^+$($\lambda$4686)      &0.000$\pm$0.000 && 0.001$\pm$0.000 \\
He/H                                &0.082$\pm$0.003 && 0.083$\pm$0.001 \\
$Y$                                 &0.248$\pm$0.009 && 0.248$\pm$0.004 \\ \hline
\end{tabular}

$^{\rm a}$ICF is the ionization correction factor.

\end{table}
%**************************************************************

%\subsubsection{Models of the H {\sc ii} region}

Inspection of Table \ref{t:Chem} shows that the ionization correction
factor ICF(S) in Pox 186 is high, compared to those for neon and argon.
In Fig.~\ref{fig:ICFs} we show for the sample of BCDs
the dependences on the ICF(S) of the
[O {\sc iii}]$\lambda$4959/[O {\sc ii}]$\lambda$3727 and
[O {\sc ii}]$\lambda$3727/H$\beta$ flux ratios [(a) and  (b), respectively] 
and oxygen abundance 12 + log(O/H) (c).
Pox 186 is shown by stars. Observations of other BCDs shown by filled 
circles are collected 
from Izotov \& Thuan (\cite{IT04}), Izotov, Thuan \& Lipovetsky (\cite{ITL97}),
Fricke et al. (\cite{F01}), Izotov, Chaffee \& Green (\cite{ICG01}), 
Izotov et al. (\cite{I99,I04}), Guseva et al. (\cite{G03}). 
It is seen from Fig. \ref{fig:ICFs} that Pox 186 deviates from other
BCDs and exhibits the largest ICF(S). 

In this paper we use the ICFs obtained from the ionization-bounded H {\sc ii}
region models of Stasi\'nska (\cite{S90}). The ICFs are fitted by 
Izotov et al. (\cite{ITL94,ITL97}) and Thuan et al. (\cite{til95}) as
functions of O$^+$/O and O$^{+2}$/O abundance ratios, 
where O = O$^+$ + O$^{+2}$. However, there is some evidence
that the ICFs for sulfur obtained from the Stasi\'nska (\cite{S90})' models
may not be correct. Recently, P\'erez-Montero \& D\'iaz (\cite{PD03}) 
presented long-slit
spectroscopic observations in the red and near-infrared of a sample of the
emission-line galaxies. Using these data and photoionization models obtained
with the CLOUDY code (Ferland \cite{F02}) they found that the S/O abundance 
ratios are $\sim$ 0.2 dex lower than those obtained with the 
Izotov et al. (\cite{ITL94,ITL97}) correction factors. Furthermore, 
Stasi\'nska \& Izotov (\cite{SI03}) produced a new grid of photoionized
H {\sc ii} region models which reproduce the relations between the observed
emission line fluxes for a large sample of BCDs. Based on these models
and additionally taking into account dielectronic recombination of sulfur and 
argon ions Izotov et al. (\cite{I04b}) obtained new ICFs for heavy elements.
%In Fig. \ref{fig:heavy}c are shown S/O abundance ratios derived with the old
%ICFs. 
The S/O abundance ratios for BCDs from the comparison sample
recalculated with new ICFs are decreased by $\sim$ 0.15 dex compared to the 
values shown in Fig. \ref{fig:heavy}c, in agreement with
P\'erez-Montero \& D\'iaz (\cite{PD03}). The effect is larger for Pox 186,
$\sim$ 0.25 dex. However, even with the new ICFs the S/O ratio in Pox 186 
is $\sim$ 0.2 dex higher than in other BCDs.

The H {\sc ii} region in Pox 186 is characterised by a very
high O$^{+2}$/O$^+$ abundance ratio of $\sim$ 20 (Table \ref{t:Chem}). 
Hence, almost all
oxygen in this H {\sc ii} region is in the O$^{+2}$ form. 
Such a situation may occur if the H {\sc ii} region is not ionization-bounded,
but density-bounded. 
The abundance of O$^+$ ion is sensitive to this effect because of the 
essentially same ionization potentials of H$^0$ and O$^0$. Therefore, the 
fraction O$^{+}$/O in the density-bounded H {\sc ii} region is lower than that
in the ionization-bounded H {\sc ii} region.
Then the ionization correction factors, which are
obtained for ionization-bounded H {\sc ii} regions, may not be valid for some
elements. Since the sulfur abundance is
derived from the low-ionization ions S$^+$ and S$^{+2}$ it is also sensitive
to the structure of the outer part of the H {\sc ii} region. Indeed, the 
ionization correction factor ICF(S) is proportional to O/O$^+$
in the ionization-bounded H {\sc ii} regions with low O$^+$/O abundance ratio
(Izotov et al. \cite{ITL94,ITL97,I04b}).
On the other hand, ICF(Ne) $\sim$ 1 and ICF(Ar) $\sim$ 1 in a H {\sc ii} 
region with low O$^+$/O abundance ratio, as in the case of Pox 186, 
because the abundances of neon and argon are
derived from the ions in higher ionization stages compared to sulfur. 
These ions are situated in the
inner part of the H {\sc ii} region and hence the ICFs for neon and argon 
are less sensitive to the structure of the outer zone of the H {\sc ii} region.

There is some observational evidence that the H {\sc ii} region in Pox 186
may be density-bounded. First, no H {\sc i} 21 cm emission was 
detected (Kunth et al. \cite{KMV88}).
%, which implies an upper limit 
%of 2$\times$10$^7$ $M_\odot$ to the H {\sc i} mass. 
Second, 
Corbin \& Vacca (\cite{V02}) noted that
the Ly$\alpha$ emission line is likely present in the UV spectrum of
Pox 186. 
However, this detection is not very confident because
of blending with the geocoronal Ly$\alpha$. In fact,
Ly$\alpha$ emission is only indicative of a density-bounded H {\sc ii}
region. Such emission is also expected from an ionization-bounded H {\sc ii}
region under certain conditions (e.g., Kunth et al. \cite{K03} and
references therein). 
An example is the BCD Tol 1214--277 with 
the second largest ICF(S) after Pox 186 (Fig. \ref{fig:ICFs}).
Its S/O abundance ratio is close to the mean
BCD value (e.g., Izotov et al. \cite{ICG01}).
Izotov et al. (\cite{I04}) have shown that the emission line fluxes 
in Tol 1214--277 are best reproduced by an ionization-bounded H {\sc ii}
region model.
Nevertheless, Thuan \& Izotov (\cite{ti97}) have detected the Ly$\alpha$
emission line in the UV {\sl HST} GHRS spectrum of Tol 1214--277 with the 
largest equivalent width $EW$(Ly$\alpha$) = 70\AA\ known for the BCDs.
Although there is no direct observational evidence for a
density-bounded H {\sc ii} region in Pox 186, such possibility is not excluded 
and may be helpful in understanding the high S/O abundance ratio
in this system.
It is worth noting that the detection of a weak [O {\sc i}] $\lambda$6300 
emission line suggests that some amount of clumpy neutral gas is also
present in Pox 186.

%*************************************************************
%             Fig.3
%*************************************************************
\begin{figure}[hbtp]%[tbh]
%\vspace{7.cm}
    \psfig{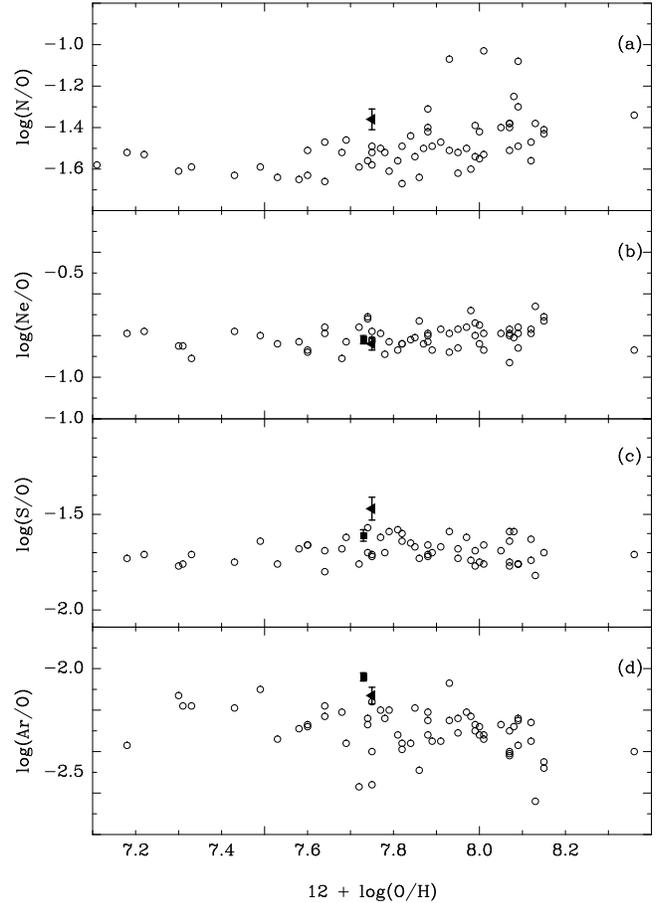}
    \caption{The comparison of the element abundance ratios in
the bright H {\sc ii} region of Pox 186 from the MMT (filled triangles) and 
3.6m data (filled squares) 
with data for other BCDs (open circles) from Izotov \& Thuan (\cite{IT99}). 
%Error bars for the heavy element-to-oxygen abundance ratios in Pox 186 
%are too small to be shown (see Table \ref{t:Chem}).  
      }
    \label{fig:heavy}
\end{figure}
%*************************************************************

%*************************************************************
%             Fig.4
%*************************************************************
\begin{figure}[hbtp]%[tbh]
%\vspace{7.cm}
    \psfig{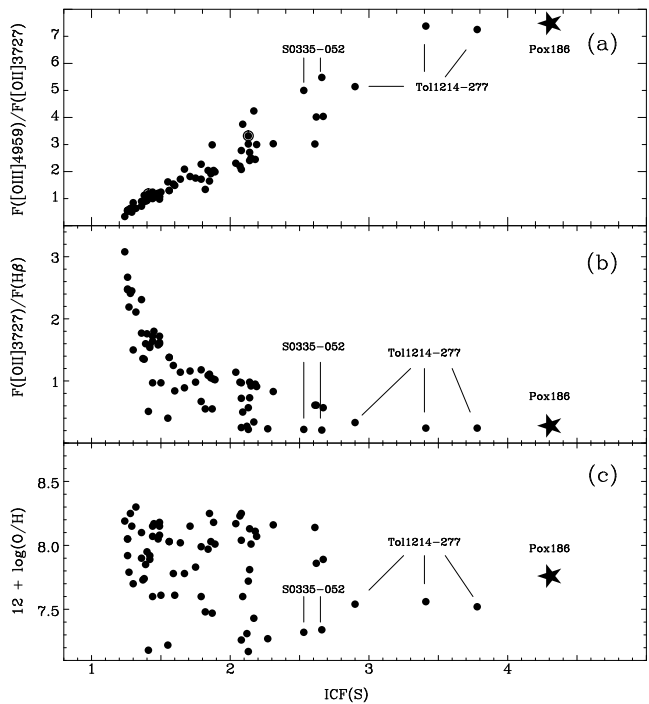}
    \caption{Emission line flux ratios
[O {\sc iii}]$\lambda$4959/[O {\sc ii}]$\lambda$3727 (a),
[O {\sc ii}]$\lambda$3727/H$\beta$ (b), and oxygen abundance 12 + log(O/H) (c)
vs ionization correction factor (ICF) for sulfur. Data for Pox 186 
(this paper) are shown by a star. 
Other BCDs are represented by filled circles (see text for references).
      }
    \label{fig:ICFs}
\end{figure}
%************************************************************

%*************************************************************
%             Fig.5
%*************************************************************
\begin{figure}[hbtp]%[tbh]
    \psfig{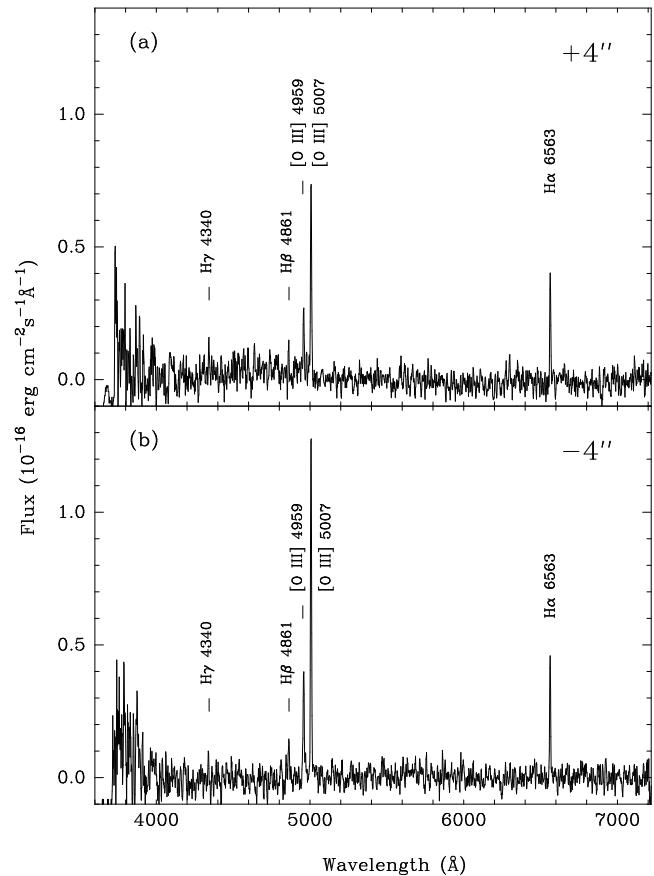}
    \caption{The spectra of Pox 186 
     at position locations --4\arcsec\ (a) and +4\arcsec\ (b)
from the center at P.A. = 0$\degr$ (MMT observations). Positive
coordinate is to the north.
      }
    \label{f:faint}
\end{figure}
%*************************************************************

%************************************************************
%                  Fig. 6
%************************************************************
%-----------------------------------------------------------------------
\begin{figure}[hbtp]
\hspace*{0.0cm}\psfig{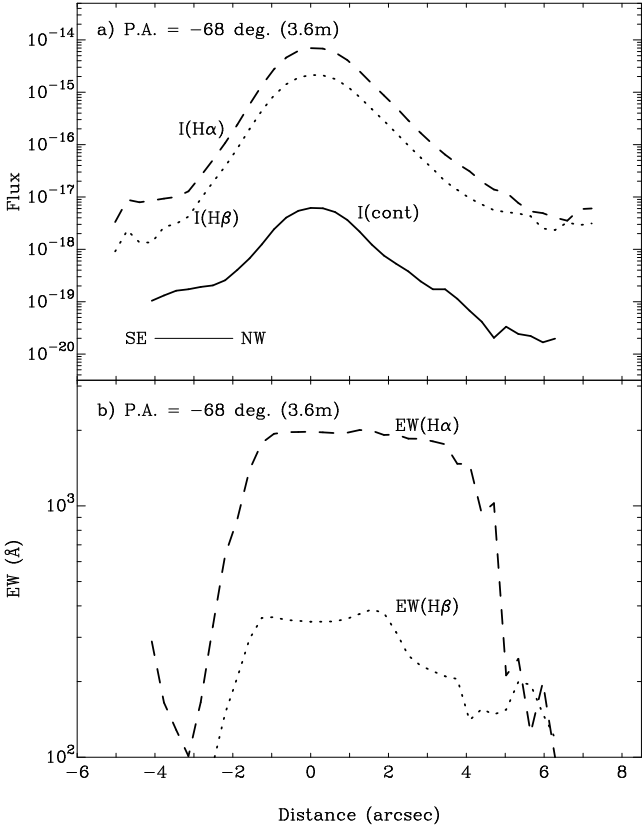}
    \caption{(a) Distributions of the fluxes of H$\alpha$
  (dashed line) and H$\beta$ (dotted line), and of the continuum near 
H$\beta$ (solid line) along the slit at P.A. = --68$\degr$ 
(3.6m observations). 
The origin is at the center of the bright H {\sc ii} region. 
 (b) Distributions of the H$\alpha$ (dashed line) and H$\beta$ (dotted line)
equivalent widths at the same position angle.
  }
    \label{f:slit}
\end{figure}
%************************************************************

\subsection{The nature and morphology of the extended LSB emission
\label{Sect:photom}}

The nature of the LSB extended emission in Pox 186 is important for 
understanding
the evolutionary status of the galaxy. Doublier et al. (\cite{D00}) and 
Corbin \& Vacca
(\cite{V02}) based on photometric data alone have concluded that the
extended emission is stellar, ruling out a significant contribution from
ionized gas. In this section we analyze the spectroscopic and photometric 
data for Pox 186 and show that, contrary to the Doublier et al. (\cite{D00})
and Corbin \& Vacca (\cite{V02}) suggestions, the emission of the
extended component is strongly contaminated by the ionized gas. 

We first consider the results of the
H$\alpha$ photometry of Pox 186 by Gil de Paz et al. (\cite{GMP03}).
From their data, the angular diameter of Pox 186 in H$\alpha$ is
$\ga$ 4\arcsec\ (or $\ga$ 360 pc in linear scale).
Gil de Paz et al. (\cite{GMP03}) have derived
an equivalent width $EW$(H$\alpha$) = 907\AA\ for the whole galaxy.
This value may be considered a lower limit because background
galaxies seen in the vicinity of Pox 186 (Fig. \ref{f:ima1})
contribute to the continuum.
Our spectroscopic data suggest that ionized gas emission
is present also at larger distances.
In Fig. \ref{f:faint} we show the spectra of Pox 186 at distances --4\arcsec\
and +4\arcsec\ from the brightest part of the galaxy along the slit at
P.A. = 0$\degr$ (MMT observations). The positive coordinates are to the
north. The strongest emission lines are clearly detected in both spectra.
% suggesting that ionized gas emission is present in the extended 
% regions at large distances.

To estimate the contribution of the ionized gas emission to the total light of 
the galaxy we obtain  the distributions along the slit at 
P.A.=--68$\degr$ (3.6m observations) of the H$\alpha$ and H$\beta$ emission 
line fluxes and equivalent widths. They are shown in Fig. \ref{f:slit}.
We also show in the figure the distribution 
of the continuum adjacent to the H$\beta$ emission line.
The H$\beta$ and H$\alpha$ emission is detected at distances out
to $\sim$ 6\arcsec\ (or 540 pc) from the center of the galaxy 
(Fig. \ref{f:slit}a).
The equivalent widths of the emission lines in
the inner region with radius $\sim$ 3\arcsec\ are very high, EW
$\sim$ 2000\AA\ in the case of H$\alpha$ emission line. Hence, the 
contribution of the gaseous continuum near H$\alpha$ is $\sim$ 50\%;
the fraction of gaseous emission is higher if H$\alpha$ is included. 
Note that the equivalent widths of the H$\beta$ and H$\alpha$ 
emission in the case of pure gaseous emission are $\sim$ 1000\AA\ and 
$\sim$ 4000\AA, respectively (Aller \cite{Aller84}).
The fraction of the gaseous continuum
near the H$\beta$ emission line in Pox 186 
is smaller, $\sim$ 30\% - 40\%. However, the 
inclusion of strong [O {\sc iii}] $\lambda$4959, 5007 emission lines results in
a larger fraction of gaseous emission in the $V$ band.
Hence, we conclude that the extended emission of Pox 186 at radii 
$\la$ 3\arcsec\ (or $\la$ 270 pc) is mainly produced by the ionized gas.
 At radii 
$\ga$ 3\arcsec\ the measurements of the equivalent widths are uncertain 
because of low continuum fluxes (Fig. \ref{f:slit}a) 
and likely contributions from background 
galaxies, although emission lines are clearly detected.

Doublier et al. (\cite{D00}) and Corbin \& Vacca
(\cite{V02}) discussed the stellar populations in Pox 186 at distances
$\la$ 3\arcsec\ where the fraction of gaseous emission to the
total light is large. Omission of gaseous emission by these authors leads
to doubtful
conclusions on the properties of stellar populations in Pox 186. 
The gaseous emission must be first subtracted to obtain
reliable results on the properties of the stellar populations in
the extended LSB component. For this the detailed determination of
the equivalent widths and relative fluxes of the nebular lines is
necessary. Additionally, much care should be exercised to take into account 
the light from the background galaxies which around Pox 186 are numerous 
(Fig. \ref{f:ima1}).

The difficulties of the analysis of stellar populations in Pox 186 are 
illustrated with Fig. \ref{f:colsects} where we show the distributions of the 
surface 
brightness in $V$ and $I$ bands and $V-I$ colour in Pox 186 at P.A. =
0$\degr$ corresponding to the location and width of the slit during the MMT 
observations. 
The origin is set to the maximum intensity in $V$. The central part
of Pox 186 is very blue with $V-I$ $\sim$ --0.8 (Fig. \ref{f:colsects}c) and
can be explained only if both stellar and gaseous emission are taken into
account. The synthetic colour of a 3 Myr old stellar population
superposed on ionized gas emission is shown by a filled circle and is in
good agreement with the observed colour. 
By contrast, a pure 3 Myr old stellar emission is by
$\sim$ 0.5 mag redder (asterisk in Fig.\ref{f:colsects}c) and is inconsistent
with observations. The colour increases with increasing distance 
along the slit and
becomes very red at distances $\sim$ 3\arcsec\ from the center. At larger
distances it gets again bluer. The very red region 
in Fig.\ref{f:colsects}c at distance between +2\arcsec\ and +4\arcsec\ is 
associated with the background galaxy seen in Fig. \ref{f:ima1} to the north
of Pox 186. It is likely that the red spike at --3\arcsec\ 
in Fig.\ref{f:colsects}c is also associated with a background galaxy.

The appreciable contribution of the ionized gas emission in Pox 186
can also be verified from the $U-V$ and $V-I$ colour maps (Fig. \ref{cmaps}). 
In this figure compact sources in the central part of the galaxy are shown 
by contours, which are computed from $V$ and
$I$ images after applying a modified unsharp-masking 
technique (Papaderos et al. \cite{P98}; Noeske et al. \cite{N03}). 
Magnified versions of these contour maps with the sources labeled 
are shown in the lower-left corners of each panel.
The  ellipses in both panels indicate the location of the 
curved feature {\it S}. 
This feature has been suggested by Corbin \& Vacca (\cite{V02}) to be tidally
ejected material in a recent collision between two smaller galaxies.

The central starburst emission splits into a pair  of compact 
clusters {\it a} and {\it b} with diameters $\leq$10 pc in $V$ and a
combined absolute $V$ magnitude, corrected for the local
background, of $M_V\sim -12$ mag. The cluster {\it b} is by $\sim$ 0.6 mag 
bluer in $V-I$ than cluster {\it a}, whereas the opposite trend is seen in 
the $U-V$ (Fig. \ref{cmaps}).
This is probably due to the growing importance of nebular line emission 
westwards of cluster {\it a}, rather than due to extinction or age effects.
This is because the $V-I$ colour in cluster {\it b}, $\la$ --1 mag, is
too blue to be accounted for by stellar emission alone. 

Regions with red $U-V$ and blue $V-I$ colours in Fig. \ref{cmaps} 
are 1) spatially correlated 
and 2) more extended (with diameters $\sim$ 200 pc) than the compact stellar 
clusters. Such a colour distribution is characteristic not of stellar emission,
but of extended ionized gas emission surrounding a more compact ionizing 
source. This is consistent with the very blue $R-I$ colour of $\sim$ --1 mag
inside a radius of 2\arcsec\ read off Fig. 1 of Doublier et al. 
(\cite{D00}).
%Hence, photometric results further support the conclusions made from 
%the spectroscopic data.
                   
%*****************************************************************
%              Fig.7
%*****************************************************************
  \begin{figure}%[hbtp]
\vspace{1.cm}
    \hspace*{-0.0cm}\psfig{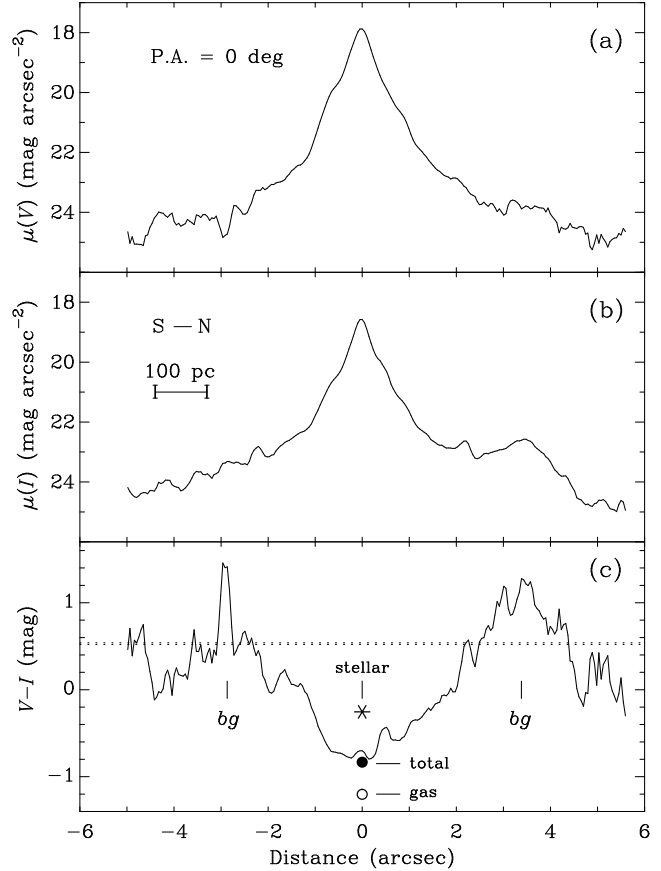}
   \caption[]{(a) and (b) {\sl HST} $V$ and $I$ surface brightness 
distributions along the slit at P.A. = 0$\degr$.
The origin is set at the intensity maximum.
(c) $(V-I)$ colour distribution. 
Locations of background galaxies are labeled {\it bg}.
    }    
\label{f:colsects}
\end{figure}
%******************************************************************

Corbin \& Vacca (\cite{V02}) have proposed that Pox 186 is the result
of the collision of two smaller galaxies.
The available photomeric data does not
allow for a decisive check of this hypothesis.
Pox 186 possesses an irregular LSB component with no 
signs of tidal tails, at least down to $\approx$25.5 $V$ mag arcsec$^{-2}$.
The colours of feature {\it S} are consistent with those for
nebular gas emission. Although the possibility of intense star
formation within a stellar tidal tail cannot be excluded, the overall 
evidence favours the idea that {\it S} is rather a starburst-driven 
gaseous shell.
%%------------------------------------------                                              %\subsection{Surface photometry}
%%------------------------------------------                                   

%*************************************************************
%             Fig.8
%*************************************************************
\begin{figure}[hbtp]%[tbh]
%\vspace{10.cm}
 \hspace*{0.0cm}\psfig{figure=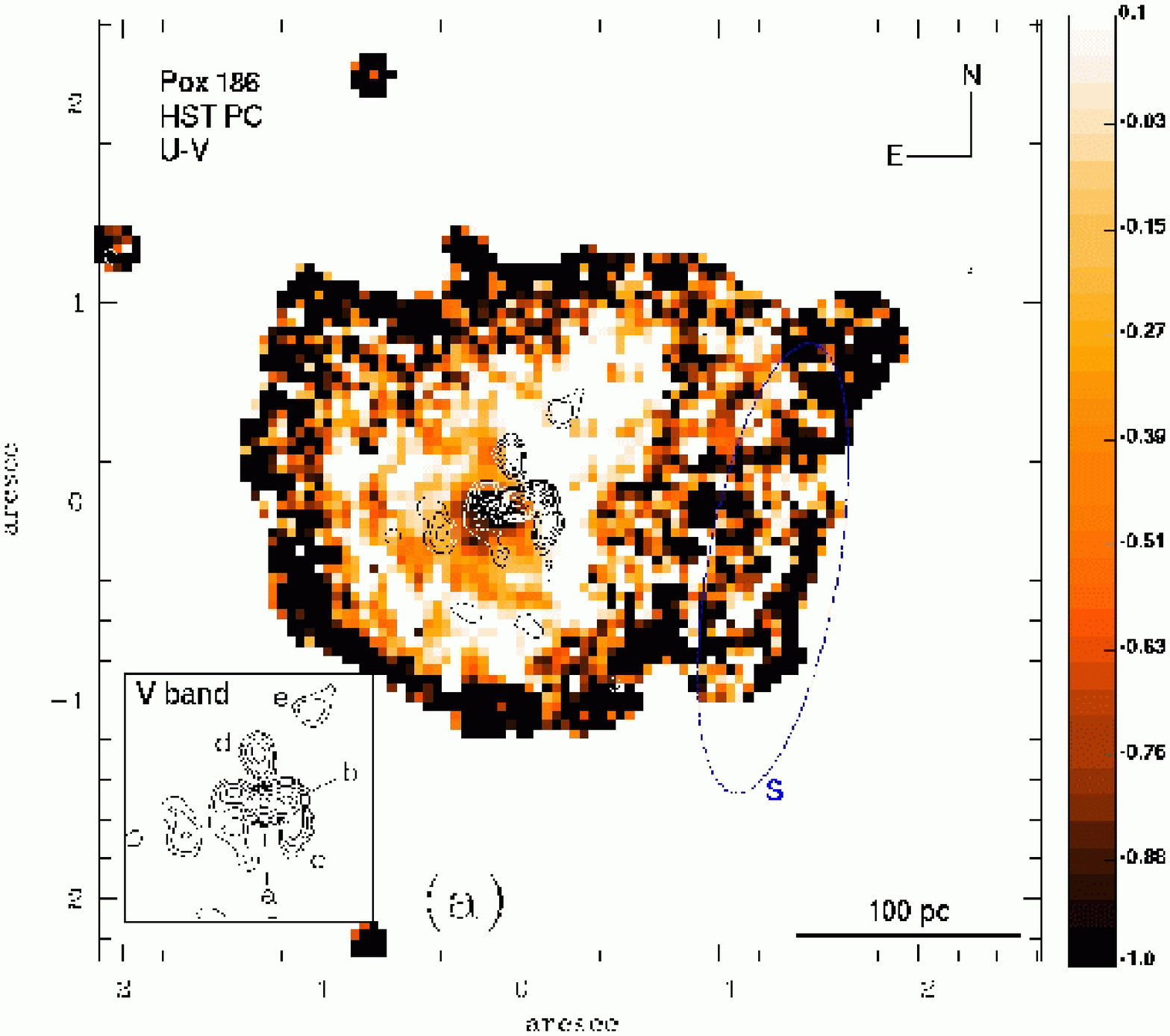,angle=0,width=8.5cm}
  \hspace*{0.0cm}\psfig{figure=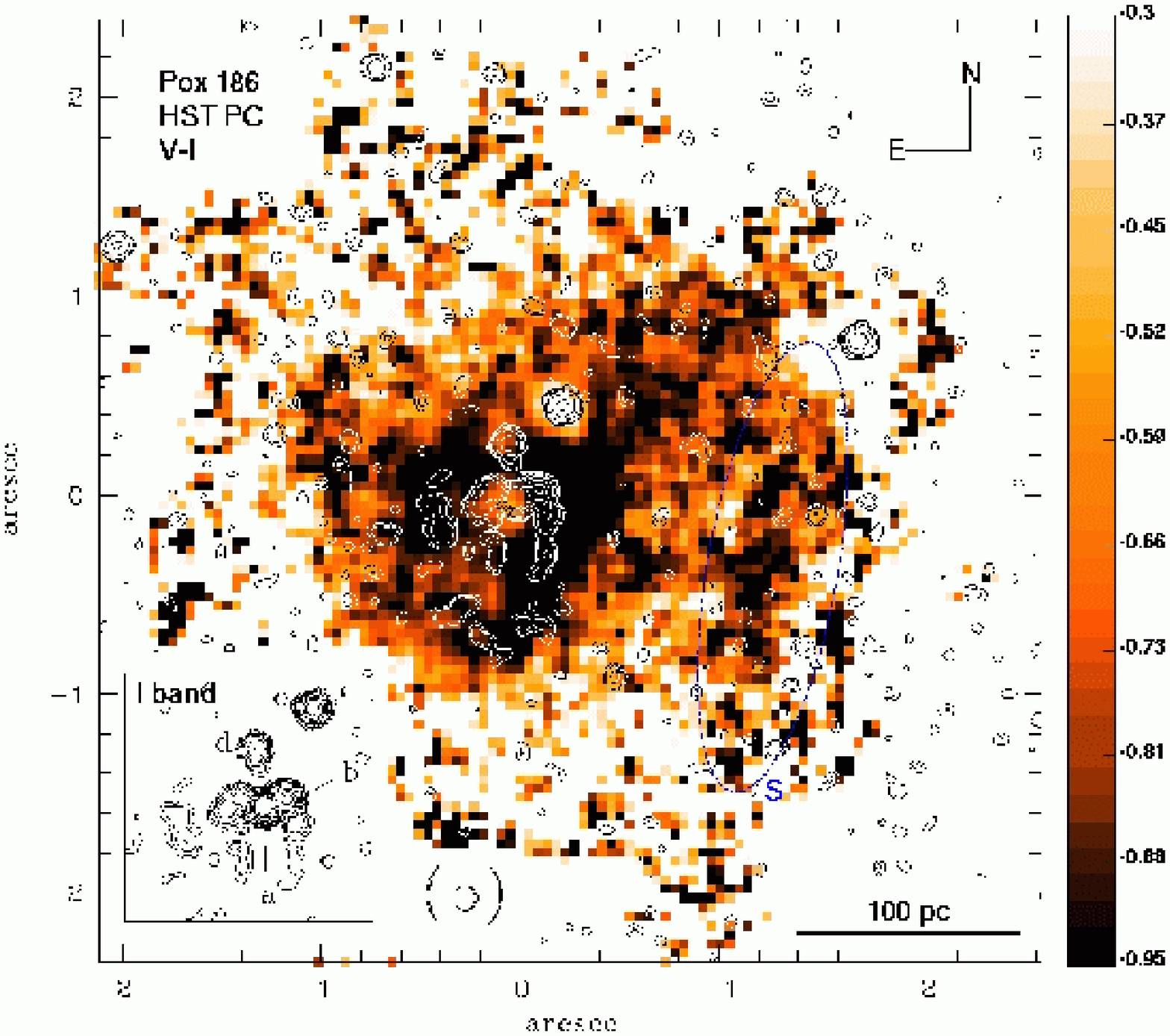,angle=0,width=8.5cm}
  \caption{(a) $U-V$ and (b) $V-I$ colour map of Pox 186
displayed in the range --1 - 0.1 mag and --0.95 - --0.3 mag, 
respectively. Compact sources detected on unsharp-masked 
$V$ and $I$ images are shown with contours.
%The sources are shown enlarged and labeled in the 
%insets at the lower-left of each panel. 
The central clusters {\it a} and {\it b} show a blue $V-I$ colour between 
--0.4 and --1.0 mag, whereas the fainter sources {\it d} and {\it e} 
display colours between --0.5 and 1.3 mag.
The position of the curved feature {\it S} (cf. Corbin \&
Vacca \cite{V02}) is marked with the ellipses. 
Note the red $U-V$ = 0.1 - 0.3 along with the blue $V-I$ = --0.8 - --1 
in the periphery of clusters {\it a} and {\it b}.
      }
    \label{cmaps}
\end{figure}
%*************************************************************

%*************************************************************
%             Fig.9
%*************************************************************
\begin{figure}[hbtp]
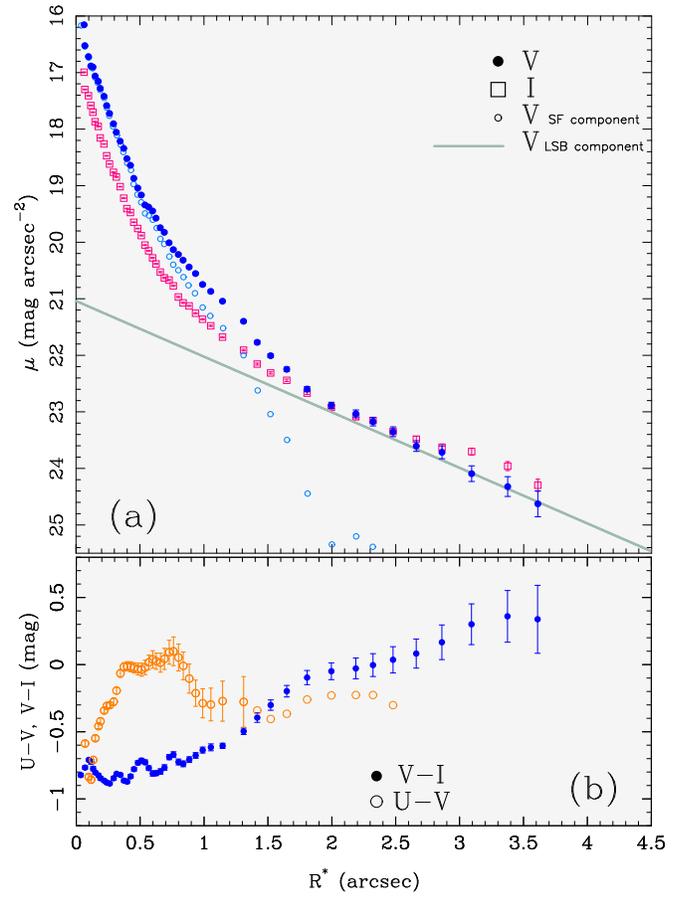
%[tbh]
\begin{picture}(16,12.0)
\put(0,4.5){{\psfig{figure=0949fig9a.ps,width=8.5cm,angle=-90.,clip=}}}
\put(0.06,0){{\psfig{figure=0949fig9b.ps,width=8.6cm,angle=-90.,clip=}}}
\end{picture}
  \caption{(a) Surface brightness profiles of Pox 186
in $V$ and $I$. The solid-grey line
is an approximation of the LSB emission by an exponential law.
The emission in excess to the exponential fit
is shown by small open circles.
%Dashed line is the ground-based $B$ band profile. 
(b) $V-I$ and $U-V$ colour profiles. 
}
    \label{sbp1}
\end{figure}

%*****************************************************************

%*****************************************************************
%              Table 4
%*****************************************************************
\begin{table} %[h]
\caption{\label{decomp_res}Decomposition of surface brightness profiles.}
\label{photom}
\begin{tabular}{lcccc}
\hline\hline
Band & $\mu_{\rm E,0}$ & $\alpha $ &  $m_{\rm LSB}$ & 
$m_{\rm SBP}$ \\
          & mag/$\sq\arcsec$ & pc       & mag &   mag \\
    (1) &   (2)            &   (3)     &  (4)      &   (5)  \\
\hline
% $B$ & 21.89$\pm$0.17  & 120$\pm$~7  & 19.27 &  17.80$\pm$0.03 \\
%
 $V$ & 21.04$\pm$1.10  & ~99$\pm$32  & 18.83 &  17.43$\pm$0.03 \\
%
% $R$ & 21.39$\pm$0.16  & 117$\pm$~7  & 18.83 &  17.65$\pm$0.03 \\
%
 $I$ & 21.26$\pm$0.44  & 120$\pm$21  & 18.63 &  18.03$\pm$0.02 \\

\hline
\end{tabular}
\end{table}
%****************************************************************

The surface brightness profiles (SBPs) of Pox 186 (Fig. \ref{sbp1}) have 
been computed using method {\sf iv} described in Papaderos et al. 
(\cite{P02}) after background sources indicated in Fig. \ref{f:ima1} have
been removed. 
Note that because the morphology of Pox 186 is not far from circular symmetry,
the equivalent radius $R^*$ of SBPs is approximately equal to the
galactocentric distance. This makes possible
an easy comparison of the distributions shown in Fig. \ref{sbp1} with
those shown in Figs. \ref{f:slit} -- \ref{cmaps} of this paper.
We decompose SBPs into a) an exponential LSB emission and 
b) the starburst emission in excess of the latter component. 
In Columns\ 2 and 3 of Table \ref{photom} we list respectively the 
central surface brightness
$\mu_{\rm E,0}$ and scale length $\alpha$ of the LSB component as obtained
from linear fits in the radius range 2\farcs9 $\leq$ $R^*$ $\leq$ 3\farcs7.
Col.\ 4 gives the total magnitude of the LSB component, extrapolated to 
infinity,
and Col. 5 the total magnitude of Pox 186, obtained from SBP integration.
Note the red $U-V$
($\approx$0) and blue $V-I$ ($\approx$--0.9) colours in the region with 
the equivalent radius in the range 0\farcs3$\leq R^*\leq$0\farcs75
(Fig. \ref{sbp1}). Such 
colours are characteristics of dominant ionized gas emission.
At larger distances $U-V$ and $V-I$ become, respectively, bluer and redder.
Such trend can partly be explained by ionized gas emission, because the 
ionization
parameter in the H {\sc ii} region decreases outwards and hence the
relative fluxes of [O {\sc iii}] $\lambda$4959, $\lambda$5007 emission lines,
contributing to the $V$ band, are also decreased. In the outermost regions with
$R^*$ $\ga$ 3\arcsec\ further reddening in $V-I$ may be caused by the faint 
background galaxies seen in Fig. \ref{f:ima1}. Because of the contribution
of ionized gas emission, photometric uncertainties and possible 
contamination by the background galaxies we do not attempt to study in detail
the properties of stellar populations in the outermost regions of Pox 186.
Nevertheless, assuming formally that the emission of ionized gas and 
background objects in the outermost regions is negligible we find that the 
$(V-I)$ colour of $\sim$ 0.4 corresponds to ages of $\sim$ 100 Myr and 
$\sim$ 1 Gyr respectively for an instantaneous burst and continuous star
formation (see, e.g., Guseva et al. \cite{G01} for details). However, these
age estimates are highly uncertain.

The comparison of $m_{\rm LSB}$ and $m_{\rm SBP}$ in Table \ref{photom} 
indicates that the starburst emission makes up at least 3/4 of the total $V$
luminosity.
The exponential scale length $\alpha$ of the LSB component was 
derived to be $\sim$ 100 pc
in $V$ and $\sim$ 120 pc in $I$.
We note, however, that both $m_{\rm LSB}$ and $\alpha$ may be significantly
affected by extended ionized gas emission which, as pointed out in 
Sect. \ref{Sect:photom}}, is present even beyond $R^*\approx$4\arcsec.
The fact that the LSB emission shows an exponential slope does not 
rule out the hypothesis that it is partly due to ionized gas emission:
as shown empirically in Papaderos et al. (\cite{P02}), 
extended H$\alpha$ emission in several star-forming dwarf galaxies can 
also be well fitted
by an exponential law in its outermost part, on scales of kpc away from the 
starburst region.
Such arguments suggest that deeper data, corrected for the effect 
of extended nebular emission and confusion with background 
sources are clearly needed to settle the question of the nature of stellar 
populations in Pox 186.

%_____________________________________________________________________________
 \section{Conclusions \label{conc}}
%-----------------------------------------------------------------------------

We use ground-based spectroscopic and {\sl HST} photometric observations of 
the ultracompact
blue compact dwarf (BCD) galaxy Pox 186 to study the properties of its bright 
H {\sc ii} 
region and the nature of the extended low-surface-brightness (LSB) component.

Our main results are as follows:

\begin{enumerate}

\item The light of the extended component in Pox 186 at radii $\la$ 3\arcsec\
(or $\la$ 270 ps in linear scale) is dominated by ionized gas
emission. This conclusion is based on the detection of H$\beta$ and
H$\alpha$ emission lines with very high equivalent widths. In particular,
the equivalent width of H$\alpha$ is $\ga$ 2000\AA\ in this region. 
The strongest emission lines
are seen at radii as large as $\sim$ 6\arcsec\ ($\sim$ 540 pc in linear
scale). However, the faintness of the galaxy at distances $\ga$ 3\arcsec\ 
precludes a reliable determination of the equivalent widths. 
Therefore, conclusions made in some previous studies on the properties 
of stellar populations in Pox 186 and its evolutionary status 
appear not to be tenable and need to be critically revised
because no correction for the contribution of the ionized gas emission was 
done.

\item At least 3/4 of the $V$-band emission of Pox 186 originates from the
star-forming component.
The surface brightness distribution of the extended LSB component 
can be approximated by an exponential law with a scale length $\alpha$ 
$\la$ 120 pc.
Since its emission is partly gaseous in origin, $\alpha$
may be even smaller. Hence, Pox 186
is one of the most compact BCDs known.

\item Values for the oxygen abundance 12 + log(O/H) = 7.76 $\pm$ 0.02 
and 7.74 $\pm$ 0.01 
($\sim$$Z_\odot$/15) are derived in the bright H {\sc ii} region of Pox 186 
from spectra obtained, respectively, with the Multiple Mirror
Telescope (MMT) and 3.6m European Southern Observatory (ESO) 
telescope. The Ne/O and Ar/O abundance ratios are close
to the mean ratios known for BCDs. On the other hand, the S/O abundance
ratio in Pox 186 is derived to be $\sim$ 2 times larger than the mean
BCD ratio. We argue, that the sulfur abundance in Pox 186 is normal,
but the apparent elevation of the S/O ratio is due to the 
a) uncertainties in the ionization correction factor for sulfur and b)
H {\sc ii} region in Pox 186 being density-bounded. 

\item The helium mass fractions in the H {\sc ii} region are 
$Y$ = 0.248$\pm$0.009 and $Y$ = 0.248$\pm$0.004 for the MMT and 3.6m 
data, respectively, suggesting high primordial helium abundance. 

\end{enumerate}

\begin{acknowledgements}
We thank the referee, Nils Bergvall, for the valuable comments which
improved the presentation of the results.
N.G.G. and Y.I.I. have been supported by DFG grant 
436 UKR 17/22/03, by Swiss SCOPE 7UKPJ62178 grant and by grant No. 02.07/00132
of the Ukrainian fund for fundamental investigations. 
They are grateful for the hospitality of the G\"ottingen Observatory. 
Y.I.I. acknowledges the G\"ottingen Academy of Sciences
for a Gauss professorship. 
P.P. and K.J.F. acknowledge support by the Volkswagen 
Foundation under grant No. I/72919. 
\end{acknowledgements} 

%------------------------------------------------------------
{}
\end{document}